# Z-States Algebra for a Tunable Multi-Party Entanglement-Distillation Protocol

Iaakov Exman  and  Radel Ben-Av

Software Engineering Department
Jerusalem College of Engineering
POB 3566, Jerusalem, 91035, Israel
iaakov@jce.ac.il, rbenav@gmail.com

**Abstract.** W-States have achieved the status of the standard fully symmetric entangled states, for many entanglement application purposes. Z-States are a generalization of W-States that display an elegant algebra, enabling short paths to desired results. This paper describes Z-States algebra starting from neat definitions and laying down explicitly some fundamental theorems on composition and distillation, needed for applications. These theorems are synthesized into a generic tunable Entanglement-Distillation Protocol. Applications are readily developed based upon the tunable protocol. A few examples are provided to illustrate the approach generality. A concomitant graphical representation allows fast comprehension of the protocol inputs, operations and outcomes.

## 1 Introduction

Whenever one needs to illustrate standard multi-qubit entangled states, it has been customary to have recourse to W-States or to GHZ-States (GHZ standing for Greenberger, Horne and Zeilinger [12]).

W-States have N terms with N entangled qubits, where in each term there is exactly one qubit with a value |1> and all other qubits with value |0>. Thus:

$$|W(N)\rangle = (|100\ldots 0\rangle + |010\ldots 0\rangle + |001\ldots 0\rangle + \cdots + |000\ldots 1\rangle) \quad (1)$$

W-States have been shown to maximize the bipartite Entanglement Formation measure [17].

This work deals with Z-States, a generalization of W-States with more degrees of freedom, viz. they have k qubits with value |1> in each term, instead of exactly one.

We develop in this paper some fundamental results of the Z-States algebra. These refer, first of all, to the ability to compose larger Z-States from smaller ones. As a consequence, we gain the capability to distill composed Z-States involving only a determined number of local operations.

The fundamental theorems are then synthesized into a generic tunable Entanglement-Distillation Protocol. The generality of the protocol is illustrated by examples showing how one readily distillates desirable Z-States. These are further depicted by a convenient graphical representation, allowing fast comprehension of the inputs, operations and outcomes.

As a matter of terminology, we shall use *state* and *vector* interchangeably (quantum states are of course rays in the Hilbert space but we shall be careful not to be misguided by terminology freedom). Moreover, all the paper deals with real valued vectors and thus we will use real valued vector spaces[1].

In the introduction we refer to related work on W-States, entanglement distillation from various points of view and previous applications of Z-States.

---

[1] We usually omit brackets from the states, unless we feel that the notation is clearer with the brackets.

## 1.1 Related Work

The prominence of GHZ-States [12] and W-States as multi-qubit entangled states has been noted by Dur et al. [9]. In particular, they observe that W-States with N=3 retain maximally bipartite entanglement when any one of the three qubits is traced out. This property is suitably generalized to any value of N. Yan et al. [22] deal with maximally entangled states of N qubits, an interesting property of N-GHZ states. They prove that two local observables are sufficient to characterize such states. Fortescue and Lo [10] define random distillation of multiparty entangled states as conversion of such states into entangled states shared between fewer parties, where those parties are not predetermined. They discuss distillation protocols for W, GHZ and particular cases of Z-States[2]. Cui et al. [5] describe ways of converting an N-qubit W-State into maximum entanglement shared between two random parties.

Experimental entanglement distillation has been achieved by several techniques. Dong et al. [8] describe a mesoscopic distillation of deterministically prepared entangled light pulses that have undergone non-Gaussian noise. It employs linear optical components and global classical communication. Miyake and Briegel [18] propose a multipartite distillation scheme by complementary stabilizer measurements; they construct a recurrence protocol for the 3-qubit W-state (see also Cao and Yang [4]). Fujii et al. [11] devised a scalable experimental scheme to generate N-qubit W-States by using separated cavity-QED systems and linear optics, with post-selection. Initially they generate the four-qubit W state $|W4\rangle$. Then they symmetrically breed couples of $|W_N\rangle$ states into $|W_{(2N-2)}\rangle$, a particular symmetric case of our Entanglement-Distillation protocol. Z-States have been observed experimentally in the particular case of four entangled photons with two excitations (cf. Kiesel et al. [16]). Toth [20] provides conditions to detect Z-States with N qubits, in the vicinity of N/2 qubits with value $|1\rangle$.

Entanglement characterization has been done by Stockton et al. [19] for the entanglement of symmetric states, such as GHZ and Z-States by calculations of several entanglement measures. Among the latter they obtain values for reduced entropy and entanglement of formation. Hein et al. [13] characterize the entanglement of graph states. A graph state is a special pure multi-party quantum state of a distributed quantum system. These states are mathematical graphs where the vertices of the graph are quantum spin systems and edges represent Ising interactions. Graph states are a possible generalization of the standard multi-party entangled states, say GHZ states. A more complex example is the graph associated with the quantum Fourier transform.

Distillation protocols can be designed by different approaches. Hostens et al. [15] show the equivalence of two such approaches, one based upon local operations yielding permutations of tensor products of Bell states (see also Dehaene et al. [6]) and another using stabilizer codes.

Various properties and applications of distillation are considered in the literature. Horodecki and Horodecki [14] analyse distillation of mixed states in higher dimensional systems. They provide a separability criterion for the total density matrix and its reductions. Watrous [21] considers the number of copies needed for distillation. Bombin and Martin-Delgado [3] discuss topological quantum distillation, which can be applied to dense coding and computation with magic states.

Among the recent applications of Z-States – which motivate the generic treatment of their algebra in this work – one finds an increased precision of quantum clock synchronization among multi-parties (cf. Ben-Av and Exman [2]).

## 1.2 Paper Organization

In the remainder of the paper, we introduce Z-State definitions (in section 2), explicitly state fundamental theorems on Z-States (in section 3), formulate a generic tunable Entanglement-Distillation protocol based upon the fundamental theorems (in section 4), illustrate the approach with some examples readily deduced from the protocol (in section 5), and conclude with a discussion (in section 6).

---

[2] Z-States were also coined symmetric Dicke states [5]. Dicke's paper refers to coherent radiation in a different specific context, long before the advent of quantum computation [7].

## 2 Z-States: A Generalization of W-states

Z-States are fully symmetric entangled states in which k qubits are in the state |1>. They generalize W-states in which k is restricted to k=1. In this section, we start our presentation of the Z-States algebra, with careful definitions.

### 2.1 Definitions

Definition 1 - $R_k^N$

$R_k^N$ is the subspace of states of N qubits such that exactly k qubits are in the state |1>.

The dimension of $R_k^N$ is the number of combinations $C_k^N$

$$\dim(R_k^N) = C_k^N = \binom{N}{k} = \frac{N!}{k!(N-k)!} \qquad (2)$$

For example:  $(|001101> + |101100>) \in R_3^6$

Definition 2 - $Z_k(N)$

$Z_k(N)$ is the fully symmetric state in $R_k^N$.

$$Z_k(N) = \sum_{k=\Sigma_{i=0}^N S_i} |S_1 S_2 S_3 \ldots S_N> \qquad S_i \in \{0,1\} \qquad (3)$$

In other words, $Z_k(N)$ is the sum over all states of N qubits such that a subset of k qubits are in state |1> and the rest in the state |0>.

Some immediate consequences of the above definitions are:

1. For k=0, the $Z_k$ state is not entangled : $Z_0(N) = |00000\ldots.0>$
2. For k=1, $Z_k$ states are W states : $W(N) = Z_1(N)$
3. The norm of $Z_k$ state is:

$$\|Z_k(N)\|^2 = \, <Z_k(N)|Z_k(N)> \, = C_k^N = \binom{N}{k} = \frac{N!}{k!(N-k)!} \qquad (4)$$

4. The normalized $Z_k$ state can be written:

$$\hat{Z}_k(N) = \frac{Z_k(N)}{\sqrt{C_k^N}} \qquad (5)$$

5. 0-1 symmetry : Replacing all the 0's by 1's and vice versa converts $Z_k(N)$ to $Z_{N-k}(N)$. So, the given 0-1 symmetry implies the $Z_k$ and $Z_{N-k}$ symmetry.

# 3 Z-States: Fundamental Theorems

In this section we explicitly state some Z-States' fundamental theorems. These theorems facilitate a wide variety of possible applications.

## 3.1 Composition Theorem

*Theorem 1 – Z-States Composition*

The following identity holds for given values of k and N and for any M such that $N - k \geq M \geq k$.

$$Z_k(N) = \sum_{j=0}^{k} Z_j(M) Z_{k-j}(N - M) \qquad (6)$$

Proof

As a first step it can be seen that the number of summands in the right hand side is equal to the number of the left hand side of eq. (6) due to the following combinatorial identity (see [1], section 24.1.1):

$$\binom{N}{k} = \sum_{j=0}^{k} \binom{M}{j}\binom{N-M}{k-j}$$

Now we will prove that any summand that appears in the left hand side of the equality, appears only once in the right hand side and conversely.

Any summand of $Z_k(N)$ is of the form $|S_1 S_2 S_3 \ldots S_N> \quad S_i \in \{0,1\}$. Assume one splits this element into two parts: one containing the first M qubits and the other one containing the last (N-M) qubits:

$$|S_1 S_2 S_3 \ldots S_N> = |S_1 S_2 S_3 \ldots S_M> |S_{M+1} S_{M+2} \ldots S_N>$$

Clearly there must be a unique j such that $|S_1 S_2 S_3 \ldots S_M> \in R_j^M$; it follows that $|S_{M+1} S_{M+2} \ldots S_N> \in R_{k-j}^{N-M}$. Moreover for every i: $S_i \in \{0,1\}$ thus $|S_1 S_2 S_3 \ldots S_M>$ is a summand in $Z_j(M)$ and $|S_{M+1} S_{M+2} \ldots S_N>$ is a summand in $Z_{k-j}(N - M)$ and each one appears only once in the right hand side of equation (6).

To prove the converse direction, one opens the summations and performs the multiplications in the right hand side of equation (6). Then, one easily sees that each $|S_1 S_2 S_3 \ldots S_N>$ that appears on the right hand side must be a summand of $Z_k(N)$.

**Qed**

The composition theorem is a purely algebraic result. It enables one to compose larger Z-States from smaller ones.

## 3.2 Z Distillation Theorem

The following theorem refers to the situation in which one starts from two fully symmetric entangled states, and one wishes to distill a composed state with a larger measure of entanglement. Distillation involves only local operations on 2k qubits independently of the total number of participating qubits.

*Theorem 2 – 2k-local Z Distillation*

Given two states $\psi^A = Z_k^A(N_1)$ and $\psi^B = Z_k^B(N_2)$ one can distill from them the state

$$\Psi = Z_k(N_1 + N_2 - 2k) \quad (7)$$

using local operations on k members of $\psi^A$ and k members of $\psi^B$.

Proof

The overall idea of the proof is a common and useful technique. It uses the Z-States Composition theorem twice. First, to make explicit the Z-States within A and B. Second, to collect back states after they are selected by local operations.

First let us note that the overall initial state of a system composed of states A and B is

$$|\Psi_{init}> = |\psi^A>|\psi^B>$$

$$= Z_k^A(N_1)Z_k^B(N_2) \quad (8)$$

We now use the composition theorem, equation (6), for both $Z^A$ and $Z^B$, replacing M by k.

$$|\Psi_{init}> = \sum_{j_1=0}^{k} Z_{j_1}^A(k)Z_{k-j_1}^A(N_1-k) \sum_{j_2=0}^{k} Z_{j_2}^B(k)Z_{k-j_2}^B(N_2-k)$$

$$= \sum_{j_2=0}^{k}\sum_{j_1=0}^{k} Z_{j_1}^A(k)Z_{j_2}^B(k)Z_{k-j_1}^A(N_1-k) Z_{k-j_2}^B(N_2-k) \quad (9)$$

We now replace the two summations by first summing over all the possible values of $r = j_1 + j_2$ and then by summing over all the possibilities of $j_2$ consistent with r.

$$|\Psi_{init}> = \sum_{r=0}^{2k} \sum_{j_2=\max(0,r-k)}^{\min(k,r)} Z_{r-j_2}^A(k)Z_{j_2}^B(k)Z_{k+j_2-r}^A(N_1-k) Z_{k-j_2}^B(N_2-k) \quad (10)$$

The sum over $r$ is actually a sum over $R_r^{2k}$ for the space that includes the k bits of A and the k bits of B. This way we have decomposed $|\Psi_{init}>$ to a sum over subspaces $R_r^{2k}$.

We now define the normalized $X_0$ state

$$X_0(2k) = \beta_0 \sum_{j=0}^{k} \alpha_j Z_{k-j}^A(k)Z_j^B(k) \quad (11)$$

where $\beta_0$ is the normalization factor and the $\alpha_j$'s will be shortly defined. Evidently $X_0 \in R_k^{2k}$. Thus if we project $|\Psi_{init}>$ upon $X_0$ only summands that also belong to $r=k$ will contribute.

We now perform a measurement on the k qubits of **A** together with the k qubits of **B**. We project on the axis such that one of the eigenvectors is $X_0(2k)$. In this case (i.e. when the projection result was indeed $X_0(2k)$) we get contributions only from the summands where r=k, thus, the remaining state is

$$|\Psi_{final}\rangle = \sum_{\substack{j_2=0 \\ r=k}}^{k} [\langle X_0|Z^A_{r-j_2}(k)Z^B_{j_2}(k)\rangle] \; Z^A_{k+j_2-r}(N_1-k)\, Z^B_{k-j_2}(N_2-k)$$

$$= \beta_0 \sum_{j_2=0}^{k} \langle \sum_{j_3}^{k} \alpha_{j_3} Z^A_{k-j_3}(k)Z^B_{j_3}(k) | Z^A_{k-j_2}(k)Z^B_{j_2}(k) \rangle Z^A_{j_2}(N_1-k)\, Z^B_{k-j_2}(N_2-k)$$

$$= \beta_0 \sum_{j_2=0}^{k} \alpha_{j_2} \langle Z^A_{k-j_2}(k)Z^B_{j_2}(k) | Z^A_{k-j_2}(k)Z^B_{j_2}(k) \rangle Z^A_{j_2}(N_1-k)\, Z^B_{k-j_2}(N_2-k)$$

$$= \beta_0 \sum_{j_2=0}^{k} [\alpha_{j_2} C^k_{j_2}\, C^k_{j_2}] Z^A_{j_2}(N_1-k)\, Z^B_{k-j_2}(N_2-k) \qquad (12)$$

If we choose $\alpha_j = [C^k_j\, C^k_j]^{-1}$ the final result is:

$$|\Psi_{final}\rangle = \beta_0 \sum_{j_2=0}^{k} Z^A_{j_2}(N_1-k)\, Z^B_{k-j_2}(N_2-k) = Z_k(N_1+N_2-2k) \qquad (13)$$

The last equality is again a result of the composition Theorem 1. This completes the proof of the 2k-local Z Distillation theorem.

**Qed**

## 4 Z-States Tunable Entanglement-Distillation Protocol

In this section we present a generic tunable Entanglement-Distillation Protocol for Z-States, synthesized from the theorems demonstrated in the previous section. First, it is shown as a loop containing a simple series of steps. Then, it is displayed graphically.

### 4.1 Z-States Tunable Protocol

The 2k-local Z Distillation theorem gets its name from the local operations on k members of each of the input states to the distillation process. Apparently, from equation (7), one *must* sacrifice 2k qubits from the input states, to obtain the distillation result.

In fact, the theorem is the basis of a far more flexible distillation protocol. The sources of flexibility are two parameters:

   a- *Number of distillation cycles* – distillation is done in a loop with a given number of cycles;
   b- *Number of selected qubits* – a projection operation involves a number of selected qubits that either come from input or from ancillary Z-States.

The generic Tunable Entanglement-Distillation Protocol for Z-States is seen in the next box:

> Z-States **Tunable Entanglement-Distillation Protocol**
>
> **Set** *Number-of-Distillation-Cycles* = to obtain desired Z-States;
> **Distillation-Loop**: until *Number-of-Distillation-Cycles* {
>
> 1- *Preparation* –
>    a. Choose input Z-States;
>    b. Locally prepare ancillary Z-States, if needed;
>    c. Select required qubits within the input Z-States, for projection;
> 2- *Projection* – on suitable axes, consuming extra qubits;}

The protocol, like the kernel of the proof of the Distillation theorem, essentially consists of two phases – viz. preparation and projection – that are repeated as long as needed to obtain the desired Z-States, while consuming extra qubits.

**4.2 Graphical Representation**

To appreciate the general look and feel of the Entanglement-Distillation protocol, it is represented graphically in Figure 1, for the straightforward single-cycle case of the Distillation Theorem.

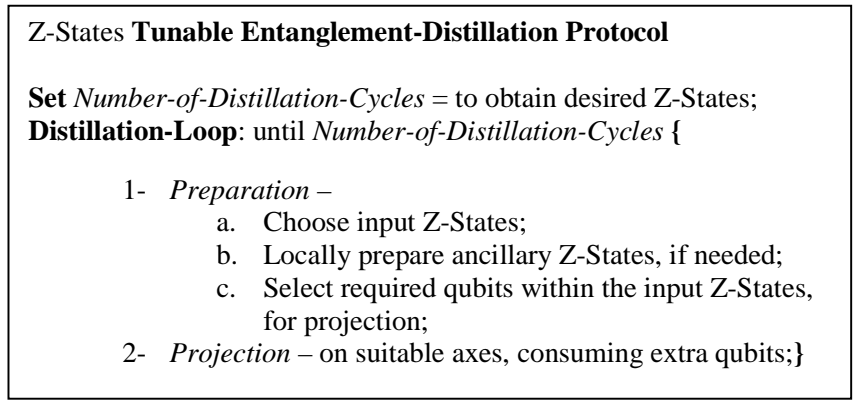

**Figure 1.** – Single-cycle Entanglement-Distillation Protocol – A Z-State distillation: preparation of the input A and B states followed by a projection consuming 2k qubits.

Z-States are graphically represented by round-rectangles. Preparation is implicitly represented by arrows selecting the suitable amount of qubits for the next operation. Projection – consuming qubits – is graphically represented by an arrow-like polygon.

Figure 1 represents graphically the fact that k qubits from each original states $Z_k^A(N_1)$ and $Z_k^B(N_2)$ are projected on $X_0(2k)$. The projection distills the original two states into a single $Z_k^A(N_1 + N_2 - 2k)$ state. One visualizes that the projection step "consumes" 2k qubits and thus the resulting number of qubits is the original number minus 2k.

# 5 Z-States: Application Examples

Here we provide examples to illustrate how one can tune the protocol parameters to readily distill desired Z-States for various applications.

## 5.1 Z-States Exact Distillation

Using the Entanglement-Distillation protocol it is possible to construct an exact – lossless – distillation that distills a $Z_k(N_1+N_2)$ state from $Z_k(N_1)$ and $Z_k(N_2)$. This is a two-cycle process. In the first cycle one uses 4k ancillary qubits locally generated. The added 4k qubits are consumed by two projection steps.

The distillation process in the first cycle includes an initial preparation of the ancillary $Z_k(4k)$ together with the original $Z_k(N_1)$, needed for a first projection step, to distill an intermediate $Z_k(N_1+2k)$. In the second cycle, the intermediate $Z_k(N_1+2k)$ is prepared with the original $Z_k^B(N_2)$, using a second projection step to create the lossless $Z_k^f(N_1+N_2)$, as described in Fig. 2.

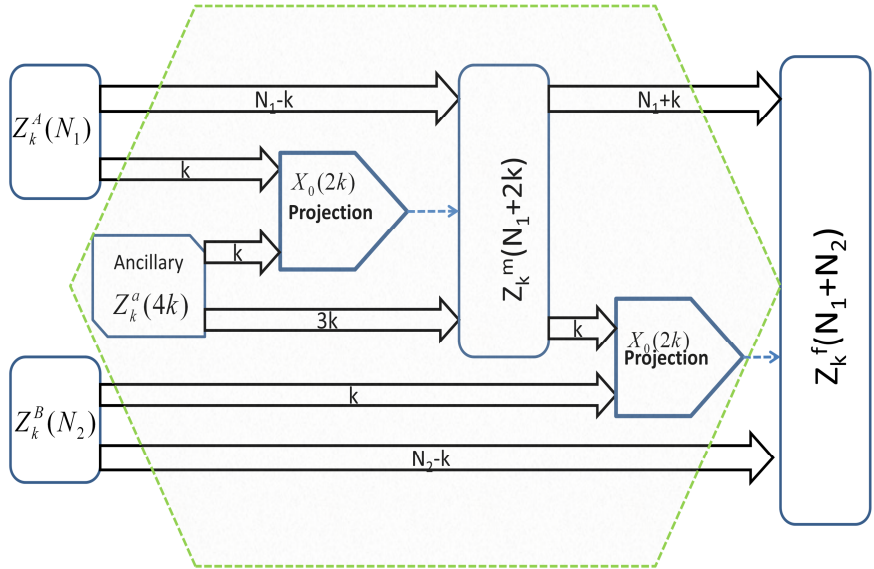

**Figure 2**. – Two-cycle exact (lossless) Z state distillation – In the first Cycle, preparation involves A together with 4k-ancilla qubits. Projection obtains an intermediate $Z_k(N_1+2k)$ state. The latter is prepared together with B in the next cycle, with projection finally obtaining the desired $Z_k(N_1+N_2)$.

Notice that the intermediate steps (inside the dashed hexagon) need not have access to all the qubits of the original states. Rather they access at most k qubits from the first state and k qubits from the second state. In addition they access additional qubits that are generated and used inside the distillation process. Thus this distillation process is also 2k-qubit local.

## 5.2 Incremental Manufacturing

In incremental manufacturing one adds qubits in an *arithmetic* progression. One starts with a supply of $Z_k(2k+1)$ states and employs only 2k-qubits operations in the following process:

1. Distill from two different $Z_k(2k+1)$ states a $Z_k(2k+1+2k+1-2k)= Z_k(2k+2)$.
2. Loop again in the same protocol – distill from a $Z_k(2k+1)$ and $Z_k(2k+2)$ states a $Z_k(2k+3)$
3. Loop again in the same protocol – distill from a $Z_k(2k+1)$ and $Z_k(2k+3)$ states a $Z_k(2k+4)$
4. And so forth

Thus, one can distill $Z_k(N)$ states for any large N using as input only $Z_k(2k+1)$ states and local projections for 2k qubits.

Now we give specific examples for W-States. The Z Distillation theorem for k=1 means that starting from two states $Z_1^A(N_1)$ and $Z_1^B(N_2)$ one can distill $Z_1(N_1 + N_2 - 2)$. Translating this to a W-States language – starting from two states $W^A(N_1)$ and $W^B(N_2)$ one can distill $W(N_1 + N_2 - 2)$, e.g. one may start with a supply of W(3) states and distill W(4). In the next loop, from W(3) and W(4) states one distills W(5). From W(3) and W(5) one distills W(6), and so on.

## 5.3 Exponential Manufacturing

In exponential manufacturing one adds qubits in a *geometric* progression. One starts with a supply of $Z_k(2k+1)$ states and employs only 2k-qubit operations in the following process:

1. Distill from two different $Z_k(2k+1)$ states a $Z_k(2k+2)$
2. Loop again in the same protocol – distill from two $Z_k(2k+2)$ states a $Z_k(2k+4)$
3. Loop again in the same protocol – distill from two $Z_k(2k+4)$ states a $Z_k(2k+8)$
4. And so forth

That is, one can distill $Z_k(N)$ states for large N using as input only $Z_k(2k+1)$ states and local projections for 2k qubits.

Again, for W-States, one may start with a supply of W(3) states and distill W(4). In the next iteration, from two W(4) states one distills W(6). From two W(6) states one distills W(10), and so on.

Of course, using exponential manufacturing takes a logarithmic number of steps relative to incremental manufacturing. However it is clear that it is possible to mix steps from both types, i.e. one can use the exponential manufacturing steps in conjunction with the incremental steps for "fine-tuning" or other reasons.

## 6 Discussion

This paper has presented and proved fundamental results of an algebra for Z-States, which are a generalization of W-States. This algebra has an interesting structure, which is revealed when the space of all states of N qubits is decomposed into subspaces, each with only k qubits in the state |1>.

These results have been synthesized into a generic tunable Entanglement-Distillation protocol. The striking simplicity of the protocol is elusive. It is a compact representation of a flexible technique that can be readily used to develop a variety of applications, as seen by the given examples.

### 6.1 Future Work

The following Z-States' issues deserve further investigation:

1. Can the distillation rate be optimized?
2. Can one achieve the desired distillation with smaller numbers of qubits from each group?
3. What is the best distillation rate for $Z_k(N)$ from the basic building blocks ($Z_k(2k+1)$)?

4. Have anti-symmetric states different properties?
5. What are suitable choices of projection axes?

A related area of interest is the Z-State distillation process in the presence of noise. We also plan to show the effectiveness of the Z-State algebra for defining and calculating entanglement measures.

There are many points of our discussion waiting to be generalized beyond Z-States. For instance:

1. Similar algebras for different sets of states;
2. Generalizations to graph states.

## 6.2 Main Contribution

The main contribution of this paper is the neat Z-States' algebra with 2k-local distillation. Its theorems are synthesized into a tunable Entanglement-Distillation protocol, from which one readily obtains desired results for a variety of applications.